\documentclass[
superscriptaddress,showpacs,floatfix,12pt,notitlepage]{revtex4-1}
\usepackage{graphicx}
\usepackage{amsmath,amssymb}
\usepackage{hyperref}
\usepackage{subfigure}

\usepackage{amsmath,amssymb}
\usepackage{braket}
\usepackage{hyperref} 
\usepackage{subfigure}
\usepackage{bm} 
\usepackage{mathtools} 
\usepackage{float}
\usepackage{cprotect}
\usepackage{setspace} 

\def\Re{{\rm Re\,}}
\def\Im{{\rm Im\,}}

\def\p{\partial}
\DeclareMathOperator{\sgn}{sgn}

\newif\iffigures
\figurestrue

\begin{document}


\title{Microscopic tunneling model of Nb-AlN-NbN Josephson flux-flow oscillator}
 


\author{D. R. Gulevich}
\affiliation{ITMO University, St. Petersburg 197101, Russia}

\author{L. V. Filippenko}
\affiliation{Kotel’nikov Institute of Radio Engineering and Electronics, Russian Academy of Science, Moscow, 125009, Russia}

\author{V. P. Koshelets}
\affiliation{Kotel’nikov Institute of Radio Engineering and Electronics, Russian Academy of Science, Moscow, 125009, Russia}


\begin{abstract}
Since the very first experimental realization of Josephson flux-flow oscillator (FFO), its theoretical description has been limited by the phenomenological perturbed sine-Gordon equation (PSGE). 
While PSGE can qualitatively describe the topological excitations in Josephson junctions that are sine-Gordon solitons or fluxons, it is unable to capture essential physical phenomena of a realistic system such as the coupling between tunnel currents and electromagnetic radiation. Furthermore, PSGE neglects any dependence on energy gaps of superconductors and makes no distinction between symmetric and asymmetric junctions:
those made of two identical or two different superconducting materials.
It was not until recently when it became possible to calculate properties of FFO by taking into account information about energy gaps of superconductors [D. R. Gulevich et al., Phys. Rev. B 96, 024215 (2017)].
Such approach is based on the microscopic tunneling theory and has been shown to describe essential features of symmetric Nb-AlOx-Nb junctions. Here we extend this approach to asymmetric Nb-AlN-NbN junctions and compare the calculated current-voltage characteristics to our experimental results.
%
%
\end{abstract}

\maketitle

\section{Introduction}
\label{intro}

Nb-based tunnel junctions are basic elements in most of the devices and circuits of low-temperature superconducting electronics~\cite{Josephson-book}. Nb-$\rm AlO_x$-Nb junctions are successfully used in SQUIDS~\cite{Clarke-v1,Clarke-v2,Seidel-2015}, RSFQ (Rapid Single Flux Quantum) digital circuits~\cite{rsfq-1,rsfq-2} and quantum computing~\cite{qc-1,qc-2,qc-3}. 
Because the noise temperature of SIS mixers is limited only by the fundamental quantum value $hf/2k_B$~\cite{Tucker-1979,Tucker-1985,Karpov-1995,Baryshev-2015},
superconductor-insulator-superconductor (SIS) mixers that employ high quality Nb-based tunnel junctions are used in the most advanced space and ground-based mm- and submm-range radio telescopes. The widespread use of the Nb-$\rm AlO_x$-Nb tunnel junctions is due to the fact that a very thin Al layer can completely cover the base Nb electrode~\cite{Rowell-1981,Gurvitch-1983,Huggins-1985} while compensating the surface roughness of the Nb film,
yielding a very high-quality tunnel barrier.

To realize a quantum-limited performance at frequencies of about 1 THz, SIS tunnel junctions with high current density, high energy gap and extremely small leakage currents are required. However, there exists a limit for increasing of the $\rm AlO_x$ barrier transparency: at values of the current density higher than about 10-15 $\rm kA/cm^2$ an abrupt degradation of the junction quality takes place. The idea of utilizing SIS tunnel junctions for heterodyne mixing at THz frequencies has received a remarkable support due to the development of Nb-AlN-Nb tunnel junctions with very high current densities up to 100 $\rm kA/cm^2$~\cite{highJ-1,highJ-2,highJ-3,highJ-4,highJ-5,highJ-6}. This corresponds to low $R_N S$ values down to 2 $\rm \Omega\mu m^2$ ($R_N$ and $S$ are the normal-state resistance and area of the SIS junction, respectively). 
Implementation of the AlN tunnel barrier in combination with a NbN top superconducting electrode provides a significant improvement in the quality of the SIS junction at high current density~\cite{highJ-6,Nb-NbN-FFOs}. In this case, the ratio of subgap to normal state resistance ($R_J/R_N$), which characterize quality of the tunnel barrier, becomes substantially enhanced. The $R_J/R_N$ as high as 28 was realized for Nb-AlN-NbN junctions at tunnel current density 20 $\rm kA/cm^2$; this value far exceeds figures for Nb-AlN-Nb ($R_J/R_N =16$) and Nb-$\rm AlO_x$-Nb ($R_J/R_N = 7$) junctions, at the same current density~\cite{Nb-NbN-FFOs}. Along with low leakage current, the Nb-AlN-NbN junctions provide high energy gap voltage~$V_g$ up to 3.7~mV, which extends considerably the operation range of SIS mixers at frequencies of around 1 THz~\cite{Khudchenko-2016}. 

High-quality Nb-AlN-NbN tunnel junctions were successfully used for development of 
Josephson flux-flow oscillator (FFO)~\cite{Nagatsuma1} which serves as a local oscillator in fully integrated superconducting receivers~\cite{Lange-2010,Koshelets-IEEE-2015}. Higher gap voltage of Nb-AlN-NbN junctions as compared to Nb-$\rm AlO_x$-Nb, results in higher Josephson self-coupling voltage $V_{\rm JSC} = V_g/3$~\cite{sc-Koshelets}
(in frequency units 620 GHz for the Nb-AlN-NbN junctions versus 450 GHz for Nb-$\rm AlO_x$-Nb), 
which provides an opportunity to engineer junction properties to suit the imposed requirements to a local oscillator.

 Despite the success in fabrication and use of Nb-AlN-NbN FFO, theory of these systems remains far from being developed. Most of the theoretical studies of FFO so far were based on the perturbed sine-Gordon equation (PSGE) which does not use any information about the material and, therefore, unable to provide an adequate description of realistic devices.
This paper aims to fill this gap by providing theoretical description of Nb-AlN-NbN FFO from the perspective of the microscopic tunneling theory (MTT)~\cite{Werthamer,Larkin}.

\section{Microscopic Tunneling Model of FFO}



FFO was proposed in 1983~\cite{Nagatsuma1} and it took years of research before a 
practical system was developed~\cite{SIR}.
Reliability of FFO as a local oscillator in high-resolution heterodyne spectrometers~\cite{Koshelets-Shitov} has been verified in field, in studies of the Earth atmosphere~\cite{Lange-2010,SIR-TELIS}, and in the lab, in measurements of radiation emitted from BSCCO intrinsic Josephson junction stacks~\cite{Koshelets-IEEE-2015,Li-2012}.
Despite the FFO has been a subject of many theoretical studies~\cite{Soriano,Golubov-FFO,Ustinov-chaos,Kurin-FFO,Cirillo,Salerno-FFO,Jaworski-1999,Yulin-FFO,Pankratov-2002-line,Pankratov-2002-driving,Sobolev-2006,Pankratov-2007,Pankratov-JAP-2007,Khapaev-2008,Pankratov-PRB-2008,Pankratov-2008,Jaworski-2010,Matrozova-2011,Revin-2012,FFO-MCQTN}, its description has been largely limited by the PSGE. The PSGE is a phenomenological theory whose treatment
of the superconducting and quasiparticle tunnel currents is only justified in a static (i.e. low-frequency) limit and close to the critical temperature~\cite{Likharev}, conditions which are rarely satisfied in practical systems. 
Furthermore, PSGE does not use any information about energy gaps of the constituting materials which is essential for systems operating in high-frequency regime.
Recently, we have initiated an approach to FFO based on the microscopic tunneling theory
and applied it to the
 description of Nb-AlOx-Nb junctions~\cite{DRGulevich-PRB-2017}.
The study yielded development of the MiTMoJCo code (Microscopic Tunneling Model for Josephson Contacts)~\cite{mitmojco} to aid calculations which use microscopic tunneling theory~\cite{Werthamer,Larkin}.
Below, we will extend this approach to a more general case of asymmetric junctions made of different superconducting materials, such as Nb-AlN-NbN.




In the study of Nb-$\rm AlO_x$-Nb FFO~\cite{DRGulevich-PRB-2017} it was shown that coupling to the SIS detector makes little or no effect to the shape of current-voltage characteristics (IVC) of FFOs. As in this paper we are mainly interested in the effect of finite superconducting energy gaps of two superconductors on the shape of IVC, here we neglect the contribution of the load and assume the FFO radiation end is unloaded.
Then, in normalized units, the quasi one-dimensional microscopic model of FFO of width profile $W(x)$ is~\cite{DRGulevich-PRB-2017},
%
\begin{equation}
\varphi_{tt} - \left(1+\beta\frac{\p}{\p t}\right)\varphi_{xx} 
- \frac{W'(x)}{W(x)}\left[h_{\rm ext} + \left(1+\beta\frac{\p}{\p t}\right)\varphi_{x}\right]
  + j(x,t) -\Gamma_{\rm eff}(x) =0,
\label{FFO-MM}
\end{equation}
\begin{equation}
\begin{split}
j(x,t)=
\frac{k}{\Re \tilde{j}_p(0) }
\int_0^{\infty}\Big\{ 
j_p(kt')\,\sin\left[\frac{\varphi(x,t)+\varphi(x,t-t')}{2}\right] \\
+\, j_{qp}(kt')\,\sin
\left [ \frac{\varphi(x,t)-\varphi(x,t-t')}{2}\right ] 
\Big\}\; dt'
\end{split}
\label{FFO-jbar}
\end{equation}
and the superconducting phase difference $\varphi(x,t)$ satisfies boundary conditions at the FFO's ends
\begin{equation}
\varphi_x(\pm L/2,t) = -h_{\rm ext}.
\label{FFO-bc}
\end{equation}
Here, $L$ and $W(x)$ are the normalized length and width of the junction, respectively,
$k = \omega_g/\omega_J$ is ratio of the gap and Josephson plasma frequencies, $\beta$ is the surface damping parameter
and $h_{\rm ext}$ is the normalized external magnetic field in units $j_c\lambda_J$. 
Assuming a mirror symmetry of the FFO layout along the $x$ axis, the effective bias current equals 
\begin{equation}
\Gamma_{\rm eff}(x)=\frac{2 h_{\gamma}(x)}{W(x)}
\label{Gamma}
\end{equation}
where $h_{\gamma}(x)$ is magnitude of the normalized magnetic field induced by the bias current along $x$.
 The spatial profile of the effective bias current $h_{\gamma}(x)$ is not known precisely as it depends on the electromagnetic environment in presence of all electrodes and circuitry. For a comprehensive numerical modeling it can be determined by a full electromagnetic calculation using the specialized software~\cite{Khapaev-1,Khapaev-2,Khapaev-3,Khapaev-4}. In this paper, we resort to a simple model where $h_{\gamma}(x)$ is taken constant. The justification for this is that while in a long superconducting strip the current rises towards the edges as $\sim 1/\sqrt{\Delta x}$~\cite{Rhoderick-1962}, where $\Delta x$ is the distance from the edges, in real FFO systems the width of the electrodes is normally made smaller than the FFO length to compensate for this rise. 


Real-valued time-domain kernels $j_p(\tau)$ and $j_{qp}(\tau)$ satisfy the causality condition
\begin{equation}
j_{p,qp} (\tau) = 0\quad\text{for}\quad \tau<0
\label{causality}
\end{equation}
and are connected to the complex quantities $\tilde{j}_p(\xi)$ and $\tilde{j}_{qp}(\xi)$ in the frequency domain by the inverse  Fourier transforms~\cite{DRGulevich-PRB-2017},
\begin{equation}
\begin{split}
j_{p}(\tau) & = \frac{1}{2\pi} \int_{-\infty}^\infty \tilde{j}_{p}(\xi) e^{i \xi \tau}d\xi
\\
j_{qp}(\tau) & = \frac{1}{2\pi} \int_{-\infty}^\infty \tilde{j}_{qp}(\xi) e^{-i \xi \tau}d\xi.
\label{transforms}
\end{split}
\end{equation}
Note the difference of the two different sign conventions of Fourier transforms which is kept for historical reasons (see note [54] in Ref.~\cite{DRGulevich-PRB-2017}).
As consequence of the causality~\eqref{causality}, the transformed quantities satisfy 
$\tilde{j}_{p,qp}(-\xi)=\tilde{j}_{p,qp}^*(\xi)$,
whereas their real and imaginary parts are connected by dispersion relations of the Kramers-Kronig type~\cite{Harris-1975,Zorin}.
The complex functions $\tilde{j}_p(\xi)$ and $\tilde{j}_{qp}(\xi)$ are referred to as tunnel current amplitudes (TCAs). The following three sections will be devoted to determination of TCAs for Nb-AlN-NbN junctions.

 

\section{Tunnel Current Amplitudes for Nb-AlN-NbN junction}

\iffigures
\setlength{\unitlength}{0.1in}	
\begin{figure}[t!]
	\begin{center}
\includegraphics[width=0.7\linewidth]{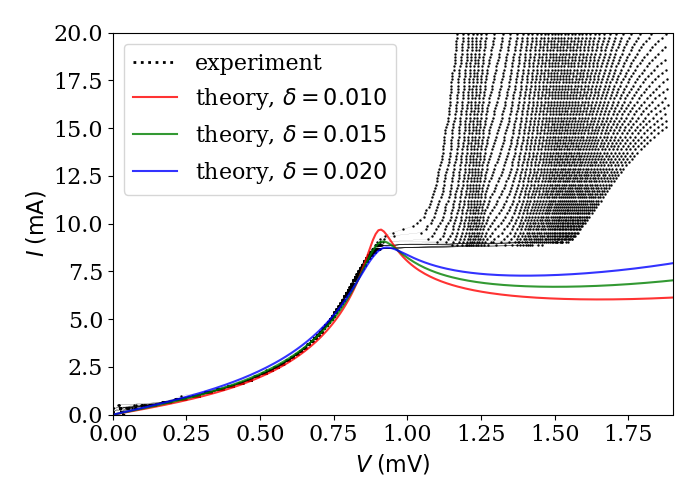}
		\caption{\label{fig:delta}
		Determination of Riedel peak smoothing parameter from the experimental IVC of FFO.
		Black dots represent experimental IVC curves of Nb-AlN-NbN FFO at high values of the magnetic field. Solid lines represent theoretical IVC curves according to the formula~\eqref{IVC-Imjqp} fitted to the experimental IVC curves at different values of the smoothness parameter $\delta$: 0.01, 0.015 and 0.02, and $V_g/R_N$ ratio $0.55\,A$.
		}		
	\end{center}		
\end{figure}
\fi

Expressions for TCAs 
were derived from the Bardeen-Cooper-Schrieffer (BCS) theory by Larkin and Ovchinnikov~\cite{Larkin}. For Josephson junction formed by superconductors with gap energies $\delta_1\equiv \Delta_1/\hbar\omega_g$ and $\delta_2\equiv \Delta_2/\hbar\omega_g$ normalized to the gap energy $\hbar\omega_g\equiv \Delta_1+\Delta_2$ and at temperature $T$ which enters via the parameter $\alpha\equiv \hbar\omega_g/2 k_B T$, these expressions are summarized in the Appendix of Ref.~\cite{DRGulevich-PRB-2017} (we refer readers to the verified expressions in Ref.~\cite{DRGulevich-PRB-2017} because the original expressions of Larkin and Ovchinnikov were given with a misprint). 
The BCS predictions differ slightly from experimental observations: in Nb junctions a smaller critical current densities are observed while the logarithmic singularities (Riedel peaks) are smeared by several competing mechanisms~\cite{Zorin,Likharev}. 
To compensate these deficiencies, phenomenological corrections are applied to the BCS results, that is 
(i) smoothing the Riedel peaks, and (ii) renormalizing the pair current density.

We correct bare BCS TCAs introducing a phenomenological peak width $2\delta$ 
and using the smoothing procedure which conserves the Kramers-Kronig transforms~\cite{Zorin}. For an asymmetric junction with $\delta_1\neq\delta_2$, assuming $0<\delta_2-\delta_1\equiv\delta_{21}$,
\begin{multline}
\Re \tilde{j}_{p,qp}(\xi) \to \Re \tilde{j}_{p,qp}(\xi)
\pm \frac{\pi \xi \sqrt{\delta_1\delta_2}}{8\delta_{21}}\left[\tanh(\alpha\delta_2)-\tanh(\alpha\delta_1)\right] \\
\times \Big[ \frac{2}{\pi}\arctan\frac{(\xi-\delta_{21})}{\delta} -\sgn(\xi-\delta_{21}) 
+ \frac{2}{\pi}\arctan\frac{(\xi+\delta_{21})}{\delta}-\sgn(\xi+\delta_{21}) \Big]
\\
-\frac{\xi\,\Re \tilde{j}_p(0)}{2\pi} 
\ln\left\{
\frac{\left[(1-\xi)^2+\delta^2\right] (1+\xi)^2}{(1-\xi)^2 \left[(1+\xi)^2+\delta^2\right]} \right\}
\label{asym-Re}
\end{multline}
\begin{multline}
\Im \tilde{j}_{p,qp}(\xi) \to \Im \tilde{j}_{p,qp}(\xi)
- \frac{\xi\sqrt{\delta_1\delta_2}}{8\delta_{21}}\left[\tanh(\alpha\delta_2)-\tanh(\alpha\delta_1)\right] \\
\times \ln\left\{ \frac{\left[(\xi-\delta_{21})^2+\delta^2\right] \left[(\xi+\delta_{21})^2+\delta^2\right]}{(\xi-\delta_{21})^2 (\xi+\delta_{21})^2} \right\} 
\pm \frac{\xi\,\Re \tilde{j}_p(0)}{2} \Big[ \frac{2}{\pi}\arctan\frac{(1-\xi)}{\delta} -\sgn(1-\xi) \\
+ \frac{2}{\pi}\arctan\frac{(1+\xi)}{\delta}-\sgn(1+\xi) \Big],
\label{asym-Im}
\end{multline}
where the plus and minus signs in Eqs.~\eqref{asym-Re},~\eqref{asym-Im} correspond to the pair and quasiparticle currents, respectively.
Value of the phenomenological smoothing parameter $\delta$ is determined by the experiment.
In Ref.~\cite{DRGulevich-PRB-2017} we used IVC of voltage biased SIS mixer made using the same technology as the FFO to determine the optimal parameter $\delta$ for smoothness of the Riedel peaks. As we will show in the next sections this parameter can also be obtained directly from the IVC of FFO. 

The pair current correction is implemented by performing a replacement
$$
\tilde{j}_p(\xi)\to \alpha_{\rm supp}\tilde{j}_p(\xi)
$$
with the suppression parameter $\alpha_{\rm supp}<1$ and leaving intact the quasiparticle component $\tilde{j}_{qp}(\xi)$. 

\section{FFO in large magnetic field limit}

Presence of a large magnetic field suppresses the Josephson effect so that the dynamics of the junction is fully determined by the quasiparticle current. 
In this regime 
theoretical description of FFO becomes particularly simple and enables 
to derive analytical formulas.

To simplify the theoretical analysis consider a FFO of constant width $W'(x)=0$ and neglect the surface damping $\beta=0$,
\begin{equation}
\varphi_{tt} - \varphi_{xx}  + j(x,t) - \gamma =0, \quad \varphi_x(\pm L/2,t) = -h_{\rm ext},
\label{MM-simplified}
\end{equation}
where $\gamma=I/I_c$ and $j(x,t)$ is given by the expression~\eqref{FFO-jbar}. Consider the limit of a very high magnetic field $h_{\rm ext}\gg 1$. In a steady state the superconducting phase difference can be taken in the form 
\begin{equation}
\varphi(x,t)\approx 2\, v_{dc} t - h_{\rm ext}x
\label{wt-hx}
\end{equation}
where $v_{dc}$ is the normalized dc voltage in units $\hbar\omega_J/e$ and the terms neglected in~\eqref{wt-hx} are of the order $O(1/h_{\rm ext}^2)$. Substituting~\eqref{wt-hx} to~\eqref{MM-simplified} and  taking the time average, we get
$$
\gamma = 
\frac{k}{\Re \tilde{j}_p(0) }
\int_0^{\infty} 
\, j_{qp}(kt')\,\sin
v_{dc} t'
\; dt'
$$
Using the causality properties of the TCAs~\eqref{causality}, we can extend the integration to the negative values of $t'$ and write
\begin{multline*}
\gamma = 
\frac{k}{\Re \tilde{j}_p(0) }
\int_{-\infty}^{\infty} 
\, j_{qp}(kt')\,\sin
v_{dc} t'
\; dt' \\
=\frac{k}{2\pi \Re \tilde{j}_p(0) }
\int_{-\infty}^{\infty}  \int_{-\infty}^{\infty}  
\, \tilde{j}_{qp}(\xi) e^{-i\xi k t'}\,\sin v_{dc} t' \, d\xi \, dt' \\
=\frac{1}{\Re \tilde{j}_p(0) }
 \int_{-\infty}^{\infty}  d\xi
\, \tilde{j}_{qp}(\xi)\,\frac{1}{2 i}\left[ \delta\left(\xi-\frac{v_{dc}}{k}\right) - \delta\left(\xi+\frac{v_{dc}}{k}\right) \right] \\
= \frac{1}{\Re \tilde{j}_p(0) }\, \Im \tilde{j}_{qp}\left(\frac{v_{dc}}{k}\right)
\end{multline*}
or, using the definition of $\gamma$ and $I_c \equiv (V_g/R_N) \Re \tilde{j}_p(0)$, in physical units,
\begin{equation}
I(V_{dc}) = \frac{V_g}{R_N}\, \Im \tilde{j}_{qp}\left(\frac{eV_{dc}}{\hbar\omega_g}\right)
\label{IVC-Imjqp}
\end{equation}
Thus, at high values magnetic fields the IVC branches converge to the imaginary part of the quasiparticle tunnel current amplitude $\Im \tilde{j}_{qp}$. 
Interestingly, the expression~\eqref{IVC-Imjqp} coincides with that for a voltage-biased small Josephson junction whose IVC is also given by $\Im \tilde{j}_{qp}$. 

\section{Determination of the Riedel peak smoothing from the IVC of FFO}
\label{sec:riedel}

The fact that the IVC branches at high magnetic field follow $\Im \tilde{j}_{qp}$
 can be used to 
extract value of the smoothing parameter from experimental data.
In Fig.~\ref{fig:delta} we present our measurements of IVC of Nb-AlN-NbN FFO sample
 at high external magnetic field values. 
Note that for $V\lesssim 0.9$ the branches condense into a single curve. 
By comparing the formula~\eqref{IVC-Imjqp} to the experimental data for FFO IVC 
we obtain an estimate for the smoothing parameter $\delta\approx 0.015$ and $V_g/R_N$ ratio 0.55 A.
In Fig.~\ref{fig:delta} theoretical dependences with $\delta=0.010$ and $\delta=0.020$ are also plotted for comparison.




\section{Numerical calculation of IVC of Nb-AlN-NbN FFO}

We use \verb|MiTMoJCo| C library~\cite{mitmojco} which implements the computationally efficient 
Odintsov-Semenov-Zorin algorithm~\cite{OSZ} and aims to assist simulations of Josephson junctions based on the MTT.
In our numerical simulations of FFO we use the TCA calculated for Nb-AlN-NbN structure assuming 1.4 meV Nb gap, 2.3 mV NbN gap, temperature $T=4.2\rm\;K$,
and the smoothing parameter $\delta=0.015$ as estimated from the experiment as described in section~\ref{sec:riedel}.
For the TCAs to be used in numerical calculations 
with \verb|MiTMoJCo|, their fits by series of exponents need to be obtained.
In our implementation of the fitting procedure we follow Ref.~\cite{DRGulevich-PRB-2017}. 
First, the desired ratio $\tau_{a}/\tau_{r}$ of the absolute $\tau_{a}$ and relative $\tau_{r}$ tolerances is chosen, which we take equal to 0.2.
Then, we fit the exact TCAs by Fourier transforms of the sum of $N$ exponentials~\cite{OSZ,DRGulevich-PRB-2017}
by minimizing the cost function
\begin{equation}
\sum_{X} \int_0^2 w\left(X^{\rm exact}\right) (X^{\rm fit}-X^{\rm exact})^2 \, d\xi
\label{cost}
\end{equation}
using the least square routine. Here, $X$ represents the functions $\Re \tilde{j}_{p}(\xi)$, $\Im \tilde{j}_{p}(\xi)$, $\Re \tilde{j}_{qp}(\xi)$, $\Im \tilde{j}_{qp}(\xi)$, respectively, and 
$w\left(X^{\rm exact}\right)$ is the weight function introduced to achieve a good fit of the TCA in the subgap region.
Unfortunately, the improper behavior of the fitted TCA in the subgap region has been a major reason of failure of early attempts to employ the Odintsov-Semenov-Zorin algorithm to description of Josephson junctions~\cite{OSZ,GJ-1992,Hattel-1993}.
Here we take the weight function $w\left(X^{\rm exact}\right)=1/\max(\tau_{a}/\tau_{r},|X^{\rm exact}|)$ which stresses 
the low-valued regions of the TCAs.
Fits of TCAs obtained using $N=8$ terms are shown in Fig.~\ref{fig:amps}a. 
The exact TCA are also shown by dashed lines, although, these are indistinguishable from the fits due to the high fit quality.
The relative errors defined as
\begin{equation}
D(X^{\rm fit},X^{\rm exact}) \equiv \frac{|X^{\rm fit}-X^{\rm exact}|}{\max(\tau_{a}/\tau_{r},|X^{\rm exact}|)}.
\label{error}
\end{equation}
are shown in Fig.~\ref{fig:amps}b. 
As seen from the figure, our TCA fits 
achieve relative tolerance $\tau_{r}=0.004$ at an absolute tolerance $\tau_{a}=0.0008$. 

Using the obtained TCA fits for Nb-AlN-NbN junction, we calculate numerically the IVC of Nb-AlN-NbN FFO 
using the MiTMoJCo code~\cite{mitmojco}.
In our numerical calculations we use parameters  
$k=4.0$, 
$\alpha_{\rm supp}=0.7$,
$\beta=0.017$, and $V_g/R_N=0.55\rm\;A$.
Our results are shown in Fig.~\ref{fig:ivcs}a. The experimental IVC is also provided for comparison in Fig.~\ref{fig:ivcs}b. Numerically calculated IVC reflects the properties of the superconducting materials 
such information about their superconducting gaps.
Similar to the experimental IVC, the numerical IVC curves at high magnetic field follows the imaginary part of the quasiparticle tunnel current according to the formula~\eqref{IVC-Imjqp}. At $V=(\Delta_{\rm NbN}-\Delta_{\rm Nb})/e$ the IVC curves exhibit a voltage step associated with the gap difference peak $\Delta_{\rm NbN}-\Delta_{\rm Nb}$ as in the experimental IVC.
The numerical IVC captures well the signatures of self-coupling: at $V_g/3$  (1.23 mV) the curves exhibit a crossover 
associated with an increase in quasiparticle current via the photon assisted tunneling. A higher order crossover associated with two photon absorption at $V_g/5$ (0.74 mV) can also be distinguished, although, this seems to be less pronounced.
There seems to be rather good agreement between the numerical and experimental IVCs in the region between 0.75 and 1.5 mV.
However, discrepancies can be observed outside this region. At smaller voltages below about 0.75 mV the 
the experimental IVC exhibits well pronounced Fiske steps which are not captured by our numerical model.
At voltages about 1.5 mV the IVC branches exhibit a cusp where the maximal flux flow current (MFFC) values reach minimum.
This turns out to be a universal feature which has been exhibited in numerical simulations of symmetric Nb-$\rm AlO_x$-Nb junctions~\cite{DRGulevich-PRB-2017}. We could not explain these discrepancies within the present model and can attribute those to a possible influence of the idle region~\cite{idle-Lee-1991,idle-Lee-1992,idle-Caputo-1994,idle-Monaco-1995,idle-Thyssen-1995,idle-Caputo-Int-1996,idle-Caputo-JAP-1999,idle-Franz-2001,idle-ZFS} which we neglect in the present model. The discrepancy for even higher voltages above 1.5 mV is associated with the influence of non-equilibrium effects which were also observed in Ref.~\cite{DRGulevich-PRB-2017} and cannot be described by the existing MTT which assumes the equilibrium distribution of quasiparticles~\cite{Larkin}.

\iffigures
\begin{figure}[h!]
\begin{center}
\includegraphics[width=\linewidth]{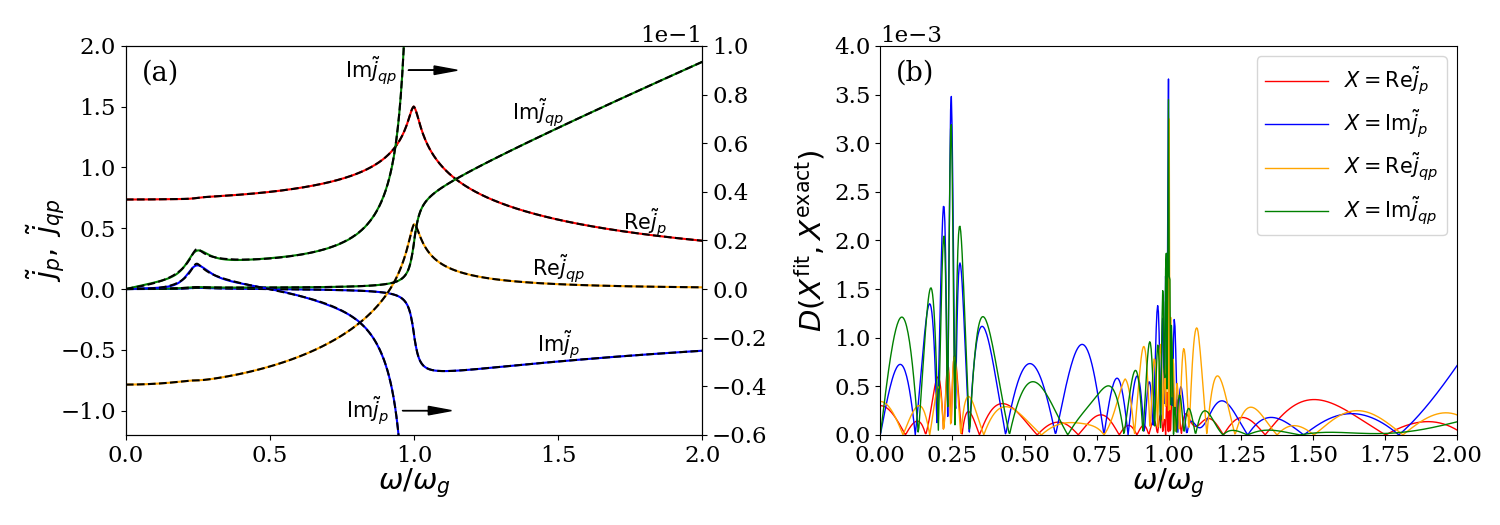}
\caption{\label{fig:amps} 
(a) Solid lines represent fits of tunnel current amplitudes (TCAs)
in the form of a sum of exponents~\cite{OSZ,DRGulevich-PRB-2017}, calculated for Nb-AlN-NbN junction assuming gap 1.4 meV for Nb and 2.3 meV for NbN, temperature $T=4.2\rm\;K$ and smoothing parameter $\delta=0.015$ estimated from the experiment, see Fig.~\ref{fig:delta}. The graphs represent TCAs after the peak smoothing procedure but {\it before} the phenomenological supercurrent suppression $\tilde{j}_{p}\to \alpha_{\rm supp}\tilde{j}_{p}$ is applied. Dashed lines are the exact tunnel current amplitudes calculated from the BCS and are indistinguishable from the fits. 
20x zoom of the imaginary parts of the tunnel current amplitudes is also shown to illustrate their behavior in the subgap region. (b) Relative differences between the fitted and exact amplitudes defined by Eq.~\eqref{error}.
}
\end{center}
\end{figure}
\fi

\section{Conclusion}

Despite a number of theoretical works dedicated to linewidth of flux-flow oscillator~\cite{Golubov-FFO,Kurin-FFO,Yulin-FFO,Pankratov-2002-line,Pankratov-JAP-2007,Pankratov-PRB-2008,Matrozova-2011,Revin-2012}, the problem of an adequate theoretical description of FFO linewidth has not been solved: the experimentally observed linewidth and existing theoretical predictions still disagree by as much as order of magnitude. Given that the previous theoretical works were based on PSGE, such disagreement should not be surprising: for reliable theoretical treatment of FFO the information about finite superconducting energy gaps of the materials should be necessarily taken into account which is ignored in PSGE.
In this paper we have introduced a theoretical description of Nb-AlN-NbN FFO based on the MTT.
Our numerical model of Nb-AlN-NbN FFO captures the features of IVC associated with finite gaps of the superconductors: self-coupling and a voltage step at the gap difference voltage $(\Delta_{\rm NbN}-\Delta_{\rm Nb})/e$. The good agreement if our numerical model with experiments raises serious expectations that it may help to solve the longstanding problem of FFO linewidth.







The presented study uncovers the intrinsic limitation of MTT itself reflected in its inability to describe effects caused by
non-equilibrium quasiparticle densities in presence of radiation with frequencies above the Nb gap frequency.
The disagreement between our experimental results and theoretical description in this frequency region can be a motivating factor for theoretical developments beyond the currently existing equilibrium MTT.

We expect that the described microscopic approach to Nb-AlN-NbN junctions will be indispensable for theoretical description of Josephson systems of a non-trivial spatial layout containing a T-junction~\cite{flux-cloning,Gulevich-2D,Gulevich-NJP,Gulevich-fluxon-pump} implemented in Nb-AlN-NbN technology.
Indeed, as has been recently shown in~\cite{DRGulevich-chaos-PRL}, presence of a T-junction may result in the appearance of the regime of chaotic self-coupling characterized by 
coupling of the tunnel currents to electromagnetic waves at all frequencies of a broad radiation spectrum rather
than the Josephson frequency exclusively.

\iffigures
\begin{figure}[t!]
	\begin{center}
	\includegraphics[width=\linewidth]{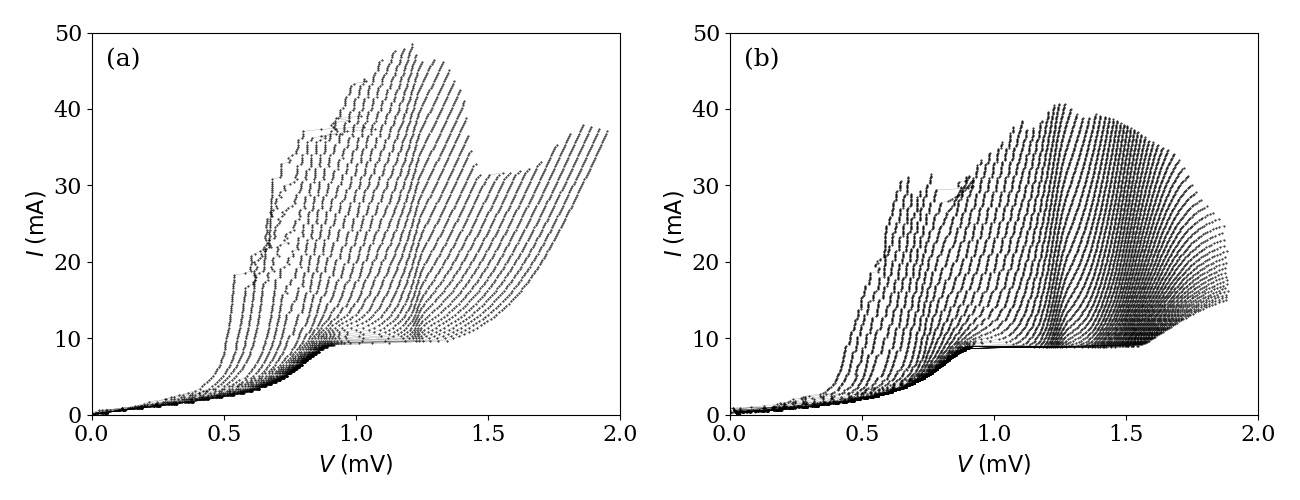}
		\caption{\label{fig:ivcs}
		(a) IVC of Nb-AlN-NbN FFO calculated numerically using the microscopic tunneling model implemented in MiTMoJCo code~\cite{mitmojco} and tunnel current amplitudes of Nb-AlN-NbN. (b) Experimental IVC of Nb-AlN-NbN FFO.
		}		
	\end{center}		
\end{figure}
\fi

\begin{acknowledgements}
The theoretical part of the work is supported by the Russian Science Foundation under the grant 18-12-00429. The experimental study is supported from the grant no.~17-52-12051 of the Russian Foundation for Basic Research.
\end{acknowledgements}




\end{document}